\documentclass{aa}  
\usepackage{graphicx}
\usepackage{txfonts}

\begin{document}

\def\kms{km s$^{-1}$}
\def\etal{et al.}
\def\hi{H\,{\sc i}}
\def\hii{H\,{\sc ii}}
\def\deg{$^\circ$}
\def\msun{M$_\odot$}
\def\mjyb{mJy beam$^{-1}$}
\def\jyb{Jy beam$^{-1}$}
\def\mdot{M$_\odot$ yr$^{-1}$}
\def\cmtres{cm$^{-3}$}
\def\cmdos{cm$^{-2}$}
\def\ojo{\fbox{\bf !`$\odot$j$\odot$!}}
\def\por{$\times$}

\title{The environs of the {H\,{\sc ii}} region Gum\,31}

\author{C. Cappa\inst{1,2}\thanks{Member of 
Carrera del Investigador, CONICET, Argentina},
V.S. Niemela\inst{1,3}\thanks{Prof. Virpi Niemela passed away on December 
18th, 2006.}, R. Amor\'{\i}n\inst{4}
\and 
J. Vasquez\inst{1,2}\thanks{Fellow of CONICET, Argentina}
}

\offprints{C. Cappa}

\institute{Facultad de Ciencias Astron\'omicas y Geof\'{\i}sicas,  Universidad
Nacional de La Plata, Paseo del Bosque s/n, 1900 La Plata, Argentina\\
\email{ccappa@fcaglp.fcaglp.unlp.edu.ar}
\and
Instituto Argentino de Radioastronom\'{\i}a, C.C. 5, 1894 Villa Elisa, 
Argentina
\and
Instituto de Astrof\'{\i}sica de La Plata, La Plata, Argentina
\and
Instituto de Astrof\'{\i}sica de Canarias, Spain\\
}
   \date{Received \today; accepted }

 
\abstract
{} 
{We analyze the distribution of the interstellar matter in the environs 
of the \hii\ region Gum\,31, excited by the open cluster NGC\,3324, 
located in the complex Carina region, with the aim of investigating the 
action of the massive stars on the surrounding neutral material.}
{We use neutral hydrogen 21-cm line data, radio continuum images at 
0.843, 2.4 and 4.9 GHz, $^{12}$CO(1-0)  observations, and IRAS and 
MSX infrared data.}
{Adopting a distance of 3 kpc for the \hii\ region and the ionizing 
cluster, we derived an electron density of 33$\pm$3 cm$^{-3}$ and 
an ionized mass of (3.3$\pm$1.1)$\times$10$^3$  M$_{\odot}$ based on the 
radio continuum data at 4.9 GHz. The \hi\ 21-cm line images revealed an 
\hi\ shell surrounding the H\,{\sc ii} region. The \hi\ 
structure is 10.0$\pm$1.7 pc in radius, has a neutral 
mass of 1500$\pm$500 M$_{\odot}$, and is expanding at 11 km\,s$^{-1}$.  
The associated molecular gas amounts to (1.1$\pm$0.5)$\times$10$^5$ 
M$_{\odot}$, being its volume density of about 350 \cmtres. This molecular 
shell could represent the remains of the cloud where the young
open cluster NGC\,3324 was born or could have originated by the shock
front associated with the \hii\ region. The difference between the ambient 
density and the electron density of the \hii\ region suggests that the 
\hii\ region is expanding. 

The distributions of the ionized and molecular material, along with that of 
the emission in the MSX band A, suggest that a photodissociation region has 
developed at the interface between the ionized and molecular gas.
The copious UV photon flux from the early type stars in NGC\,3324 keeps 
the \hii\ region ionized. 

The characteristics of a relatively large number of the IRAS, MSX, and 
2MASS point sources projected onto the molecular envelope are compatible 
with  protostellar candidates, showing the presence of active star forming 
regions. Very probably, the expansion of the \hii\ region has triggered 
stellar formation in the molecular shell.
}
   {}
   \keywords{ISM: \hii\ regions -- ISM: individual objects: 
Gum\,31  -- stars: early-type -- stars: individual: HD\,92206
               } 
\titlerunning{The environs of Gum\,31}

\maketitle
%

\section{Introduction}

The interstellar medium associated with the birth place of massive stars, 
such as O or early B-type stars, is made up of dense giant molecular clouds. 
Massive stars are characterized by intense photon fluxes which ionize and 
photodissociate the surrounding material. An \hii\ region is a direct 
consequence of the high rate of Lyman continuum luminosity. Initially, 
the \hii\ region is a small, high density region, commonly named 
ultra-compact \hii\ region (UC \hii, Wood $\&$ Churchwell 1989). If the 
photon flux rate of the massive star is sufficiently high, the \hii\ region 
evolves into a normal \hii\ region.  

A neutral shell encircles the \hii\ region during the expanding phase
(e.g. Spitzer 1978). The presence of these neutral shells is observed in 
the \hi\ 21-cm line emission distribution (e.g. Deharveng et al. 2003).
Molecular line studies have allowed the identification of molecular 
gas following the outer borders of \hii\ regions, indicating the presence 
of photodissociation regions (PDRs). Deharveng et al. (2005) have 
detected these PDRs in a number of \hii\ regions. 

In the present study, we analyze the distribution of the ionized and neutral 
material associated with the \hii\ region Gum\,31 (Gum 1955) based on \hi\ 
21-cm line emission data, radio continuum information at different 
frequencies, and IR and molecular data.

The \hii\ region Gum\,31 is about 15\arcmin\ in size and approximately 
circular in shape (Fig. 1). It is located at  {\it (l,b)} = 
(286\deg 12\arcmin, --0\deg 12\arcmin) in the complex region of Carina and 
is considered a member of the Car OB1 association. The SuperCOSMOS
image (Parker et al. 2005) shows a quite inhomogeneous \hii\ region, with 
a sharp and bright rim towards lower galactic longitudes and lower galactic 
latitudes, looking fainter and more diffuse towards higher galactic 
longitudes. 

Based on data of the radio recombination lines (RRLs) H109$\alpha$ and 
H110$\alpha$ at 5 GHz, Caswell \& Haynes (1987) found that the LSR 
velocity of the ionized gas in Gum\,31 is --18 \kms, similar to the 
velocities of other \hii\ regions in the area of the Car OB1 association 
(Georgelin et al. 1986), and derived an electron temperature $T_e$ = 7100 K.  
They estimated a flux density $S_{5GHz}$ = 35 Jy. 

The excitation sources of the \hii\ region Gum\,31 are the OB star members of 
the open cluster NGC\,3324. The brightest star in this cluster is
HD\,92206, which is the visual double star IDS\,10336-5806 in the Index
catalogue (Jeffers et al. 1963) with a 1 mag fainter companion (HD\,92206B)
placed 5\arcsec\ to the East. Another bright cluster member is located 
35\arcsec\ to the SW. This star is CD--57\deg 3378, also referred to as
HD\,92206C in the literature.

The three brightest stars in NGC\,3324 have published spectral types. Both
HD\,92206A and B are classified as O6.5V, and HD\,92206C as O9.5V by
Mathys (1988). Walborn (1982) classifies HD\,92206A as O6.5V(n) and the
component C as O8.5Vp.

Moffat $\&$ Vogt (1975) first carried out photometric observations 
of about 12 stars in the cluster and estimated a color excess $E(B-V)\sim$ 
0.45$\pm$0.05 mag and a distance $d$ = 3.3 kpc. Clari\'{a} (1977), using 
UBV photometry, confirmed previous results by Moffat $\&$ Vogt and 
estimated $d$ = 3.1 kpc. Vazquez \& Feinstein (1990) derived a distance 
$d$ = 3.6 kpc for the cluster from UBVRI photometry. More recently, 
Carraro et al. (2001) found about 25 new possible cluster members and 
derived  $d$ = 3.0$\pm$0.1 kpc. On the other hand, distance estimates for 
Car\,OB1 are in the range 1.8-2.8 kpc (Walborn 1995). Bearing in mind 
these results we adopted  $d$ = 3.0$\pm$0.5 kpc for both Gum\,31 and 
the ionizing cluster.

Carraro et al. (2001) find evidence for pre-main sequence members beginning 
at about late B spectral type, which suggests an extremely young age for 
NGC\,3324 ($\leq$ (2-3)\por 10$^6 $ yr). The O-type members
would be stars recently arrived on the Zero Age Main Sequence.

\begin{figure}
\resizebox{8.8cm}{!}{\includegraphics{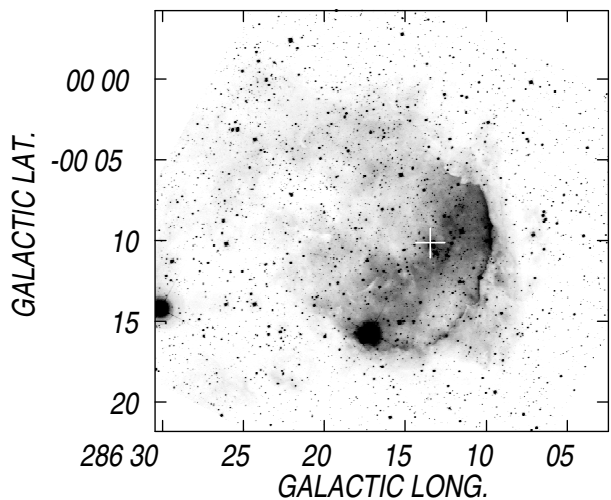}}
\caption{SuperCOSMOS image showing the \hii\ region Gum\,31. The 
 cross  marks the position of the multiple system HD\,92206. 
The intensity units are arbitrary.
 }
\label{optical image}
\end{figure}

\begin{table}
\centering
\caption[]{Radio  data: relevant parameters.}
\begin{tabular}{lc}
\hline
Radio continuum at 4.85 GHz \\
\hline
Angular resolution             &  5\farcm 0 \\
rms noise                      &  10 \mjyb \\
\hline
Radio continuum at 2.4 GHz \\
\hline
Angular resolution             &  10\farcm 4 \\
rms noise                      &  12 \mjyb \\
\hline
Radio continuum at 0.843 GHz \\
\hline
Angular resolution      &  43\arcsec\ $\times$ 51\arcsec \\
rms noise                      &  1 \mjyb \\
\hline
\hi\ data \\
\hline
Synthesized beam               &   2\farcm 4 $\times$ 2\farcm 1 \\
Number of channels             & 256 \\
Velocity coverage              &   (--190,+230) \kms \\
Velocity resolution            &   1.64  \kms  \\
RMS noise level                &  1.6 K \\
\hline
$^{12}$CO(1-0) NANTEN data \\
\hline
Angular resolution             &  2\farcm 7 \\
Velocity coverage              &  (--50,+50) \kms \\
Velocity resolution            &   0.2 \kms  \\
RMS noise level                &   1.0 K \\
\hline
\end{tabular}
\end{table}
\section{Data sets}

\subsection{Radio data sets}

We analyzed the radio continuum emission in the region of Gum\,31 using
data obtained at 0.843, 2.42 and 4.85 GHz, which were extracted 
from the Sydney University Molonglo Sky Survey (SUMSS) (see Sadler \&
Hunstead 2001 for a description), the survey by Duncan et al. (1995), and 
the Parkes-MIT-NRAO (PMN) Southern Radio Survey (see Condon et al. 1993 for 
a complete description of this survey), respectively. 

We used \hi\ data from the Southern Galactic Plane Survey (SGPS) to
analyze the neutral gas distribution in the environs of Gum\,31.
These data were obtained with the Australia Telescope Compact Array 
(ATCA) and the Parkes Radiotelescope (short spacing information).
A Hanning smoothing (Rohlfs 1986) was applied to the \hi\ data to improve 
the signal to  noise ratio. A description of this survey can be found in 
McClure-Griffiths et al. (2005).

The distribution of the molecular material in the region was studied
using  $^{12}$CO(1-0) line data at 115 GHz obtained with the NANTEN 4 m 
telescope of Nagoya University at Las Campanas Observatory of the Carnegie 
Institution of Washington, and published by Yonekura et al. (2005). 
Observational parameters for these data were taken from Yonekura et al. 
(2005).

The main observational parameters of these databases are listed in Table 1.

\subsection{Infrared data}

We also investigated the dust distribution using high-resolution
(HIRES) IRAS, and MSX data obtained through {\it IPAC}\footnote{{\it IPAC} 
is funded by NASA as part of the {\it IRAS} extended mission under 
contract to Jet Propulsion Laboratory (JPL) and California Institute of 
Technology (Caltech).}. The IR data in the {\it IRAS} bands at 60 and 
100 $\mu$m have angular resolutions of 1\farcm 1 and 1\farcm 9. 
The images in the four MSX bands (8.3, 12.1, 14.7, and 21.3 $\mu$m) have
an angular resolution of 18\farcs 3. MSX units were converted into 
MJy ster$^{-1}$ by multiplying the original figures by 7.133\por 10$^6$
(Egan et al. 1999). 

\section{Results}

\subsection{Radio continuum emission}

\begin{figure}[!hbtp]
\resizebox{8.8cm}{!}{\includegraphics{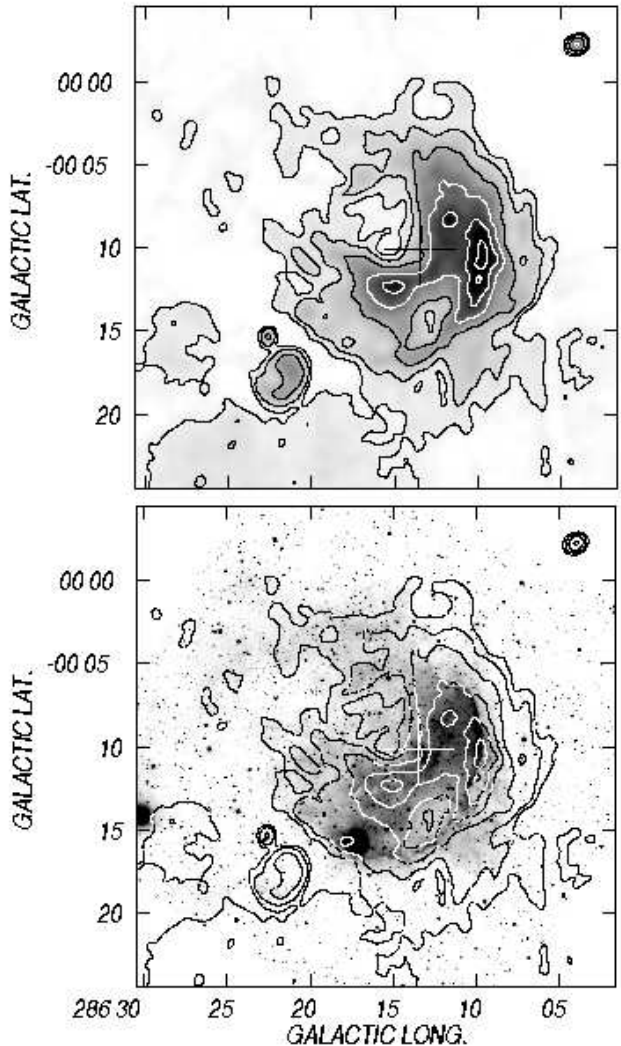}}
\caption{{\it Top:} Radio continuum image at 843 MHz. The grayscale 
is from -5 to 180 m\jyb. The contour lines are 5 (5$\sigma$), 20, 50, 100,
150, and 200 m\jyb. The cross indicates the position of HD\,92206. 
{\it Bottom}: Overlay of the image at 843 MHz (contour lines) and the
SuperCOSMOS image of Gum\,31 (grayscale).
 }
\label{}
\end{figure}

Figure 2 displays the radio continuum image at 843 MHz. The image reveals 
a radio source of 15\arcmin\ in size, coincident in position with the 
\hii\ region. There is a remarkable correlation between radio and optical 
emission regions. The strongest radio emission region coincides with the
brightest section of the  optical rim at {\it l} $\simeq$ 286\deg 
10\arcmin. The region of diffuse emission near {\it (l,b)} = 
(286\deg 20\arcmin, --0\deg 4\arcmin) also has a weak radio counterpart. 
The fainter optical regions at {\it (l,b)} = (286\deg 13\arcmin, 
--0\deg 15\arcmin) and (286\deg 20\arcmin, --0\deg 7\arcmin) also correlate
with weak radio emission. Along with the optical image, this radio 
image shows that the {H\,{\sc ii}} region is far from being homogeneous 
and that it is non-uniform in density. Gum\,31 is also detected in the 
surveys at 4.85 and 2.4 GHz as an isolated radio source. The flux 
densities at 4.85 and 2.42 GHz are 37.7$\pm$2.5 Jy (coincident with the 
previous estimate by Caswell \& Haynes [1987]) and 40.7$\pm$2.9 Jy, 
respectively. 

The spectral index of the radio source, as derived from the emissions at 
2.4 and 4.85 GHz, is $\alpha$ = --0.09$\pm$0.20, confirming its thermal 
nature. This result is compatible with the detection of RRL at 5 GHz. 

\subsection{\hi\ results}

\begin{figure*}
\resizebox{17cm}{!}{\includegraphics{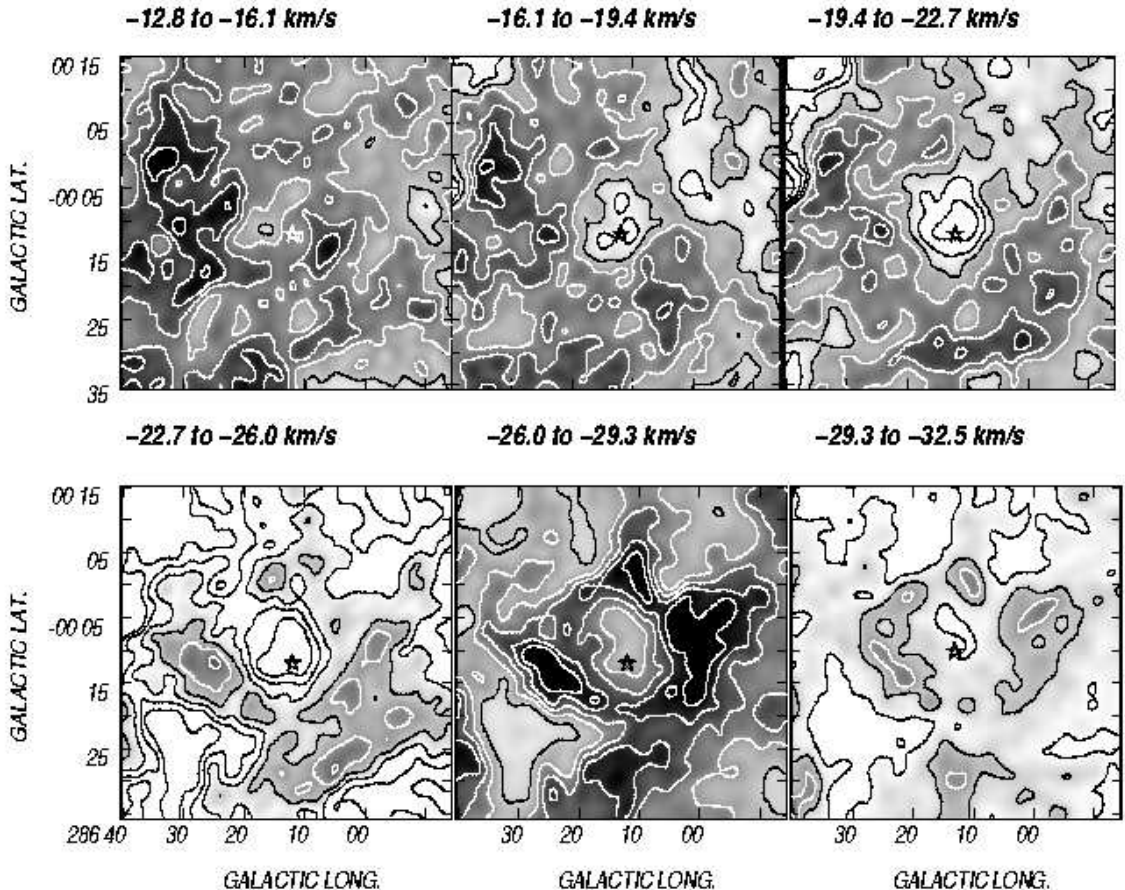}}
\caption{Series of \hi\ brightness temperature images for the
velocity interval --12.8 to --32.5 \kms\ averaged over 3.3 \kms. The greyscale
is from 70 to 120 K for the images at ${\rm v} >$ --26.0 \kms, and 
from 10 to 60 K for the images at ${\rm v} <$ --26.0 \kms. The contour 
lines are from 10 to 110 in steps of 10 K. The star indicates the position of
HD\,92206. The velocity interval corresponding to each image is indicated.}
\label{series}
\end{figure*}
 
Circular galactic rotation models predict negative radial velocities of up 
to about --12 \kms\ in the line of sight towards {\it l} = 286\deg. 
However, radial velocities observed in this section of the Galaxy are more
negative (of up to --30 \kms, Brand \& Blitz 1993), indicating the presence 
of non-circular motions. Consequently, we paid special attention to the 
neutral gas emission distribution at negative radial velocities of up to 
--50 \kms. 

Figure 3 displays a series of \hi\ images within the velocity interval 
from --12.8 to --32.5 \kms\ in steps of 3.3 \kms. The presence of a
region of low \hi\ emission surrounded by regions of enhanced emission 
approximately centered at the position of HD\,92206 is clearly identified 
within the velocity range --12.8 to --32.5 \kms. An almost complete 
envelope encircles the void.

The top panel of Fig. 4 displays the \hi\ emission distribution within 
the velocity range --29.3 to --16.5 \kms, where the relatively thick 
envelope,  of about 8\arcmin -10\arcmin\ (or 7.0-8.7 pc at $d$ = 3 kpc), 
is clearly defined. Within this velocity range, the envelope 
is more easily identified at higher negative velocities, where the 
approaching part of the shell is detected. 

The bottom panel of Fig. 4 displays an overlay of the \hi\ emission 
distribution and the SuperCOSMOS image of Gum\,31. The \hi\ envelope clearly 
anti-correlates with the ionized nebula. This envelope is
less bright towards higher galactic latitudes.

The systemic velocity of the structure, which corresponds to the velocity
at which the feature presents its largest dimensions and deepest temperature
gradient, is about --23 \kms. This velocity is similar to the
velocity of the ionized gas (--18 \kms, see Sect. 1) as obtained from RRLs,
within the errors.

The morphological agreement between the optical nebula and the \hi\  
shell and the agreement in velocity between the ionized and neutral 
materials indicate that the \hi\ feature is the neutral counterpart of the
ionized nebula.  

\begin{figure}[!hbtp]
\resizebox{8.8cm}{!}{\includegraphics{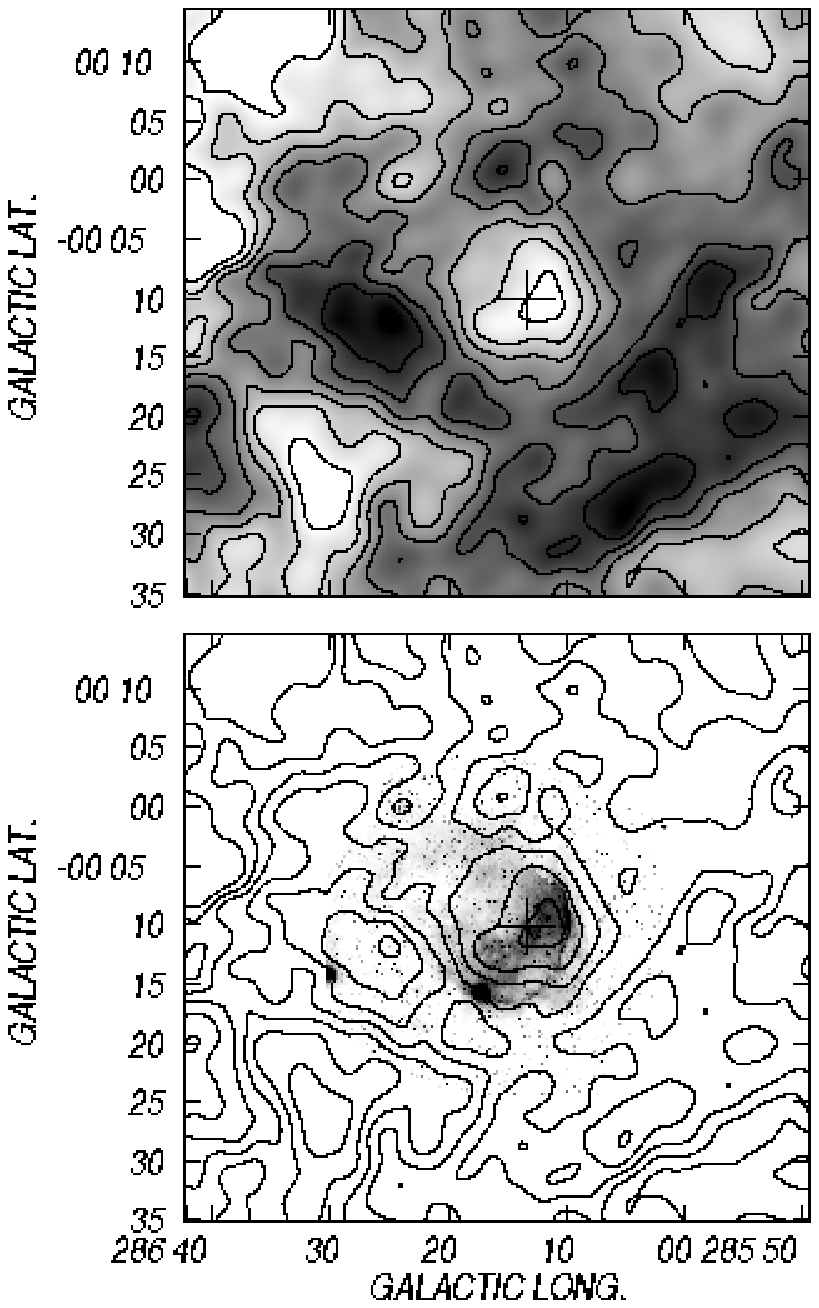}}
\caption{{\it Top:} \hi\ brightness temperature image corresponding to the
velocity interval --29.3 to --16.5 \kms\ showing the \hi\ structure 
related to Gum\,31. The grayscale is from 40 to 100 K. The contour lines are 
from 40 to 100 K in steps of 10 K.
{\it Bottom:} Overlay of the SuperCOSMOS ({\it grayscale)} and \hi\ 
({\it contours}) images showing the close correspondence of neutral and 
ionized gas emissions. 
 }
\label{nh}
\end{figure}

\subsection{The CO emission distribution}

The distribution of the molecular gas is shown in Fig. 5. The upper panel 
displays the $^{12}$CO(1-0) emission distribution in grayscale and 
contour lines, while the bottom panel displays an overlay of the CO contour 
lines and the optical image of Gum\,31.
The CO gas distribution displayed in the figure was obtained from 
the molecular data cube kindly provided by Y. Yonekura. We integrated the 
$^{12}$CO(1-0) emission within the velocity interval --27.2 to --14.0
\kms. This velocity range is slightly different from that used by Yonekura 
et al. (2005) (--30 to --10 \kms), since no CO emission was detected for 
velocities ${\rm v} <$ --27 \kms, and ${\rm v} >$ --14 \kms\  associated 
with Gum\,31. Both images, that by Yonekura et al. and the one in Fig. 5, 
are essentially the same. 

Intense CO emission regions encircle the brightest sections of the optical 
nebula with CO clumps strikingly bordering the bright rim at 
(286\deg 10\arcmin, --0\deg 10\arcmin), and 
near (286\deg 23\arcmin, --0\deg 15\arcmin), where the nebula appears 
diffuse. The CO envelope is open towards {\it (l,b)} = 
(286\deg 25\arcmin, --0\deg 5\arcmin). The region of relatively faint 
optical emission near {\it (l,b)} = (286\deg 20\arcmin, --0\deg 4\arcmin) 
coincides with a low CO emission region. 
 
The close morphological agreement between the optical and the molecular 
emissions in the brightest optical region indicates that the molecular 
material is interacting with Gum\,31. 

The thick lines in the bottom panel of Fig. 5 delineate the C$^{18}$O 
cores found by Yonekura et al. (2005). The dense cores coincide with the
the brightest $^{12}$CO emission regions.
 
The comparison of the molecular gas distribution as shown by the $^{12}$CO
emission  (Fig.~5) with the \hi\ structure (Fig.~4) shows that the neutral 
envelope around the \hii\ region has a molecular component. Although, both 
the  \hi\ and the molecular envelopes are approximately coincident, most of  
the bright molecular clumps anticorrelate with the \hi\ maxima, indicating 
that gas in these regions is mainly in molecular form. However, the 
comparison of the CO distribution with the \hi\ emission distribution at 
different velocities shows that the $^{12}$CO clump at 
{\it (l,b)} = (286\deg 23\arcmin, --0\deg 15\arcmin) coincides with
an \hi\ cloud present in the velocity interval from --26 to --29.3 \kms. 
A possible explanation is the presence of the CO clump inside 
a dense \hi\ cloud. 
  
The comparison between the ionized, neutral atomic, and molecular 
distributions around Gum\,31 suggests the presence of a prominent 
PDR bordering the brightest ionized regions. 

\begin{figure}
\resizebox{8.8cm}{!}{\includegraphics{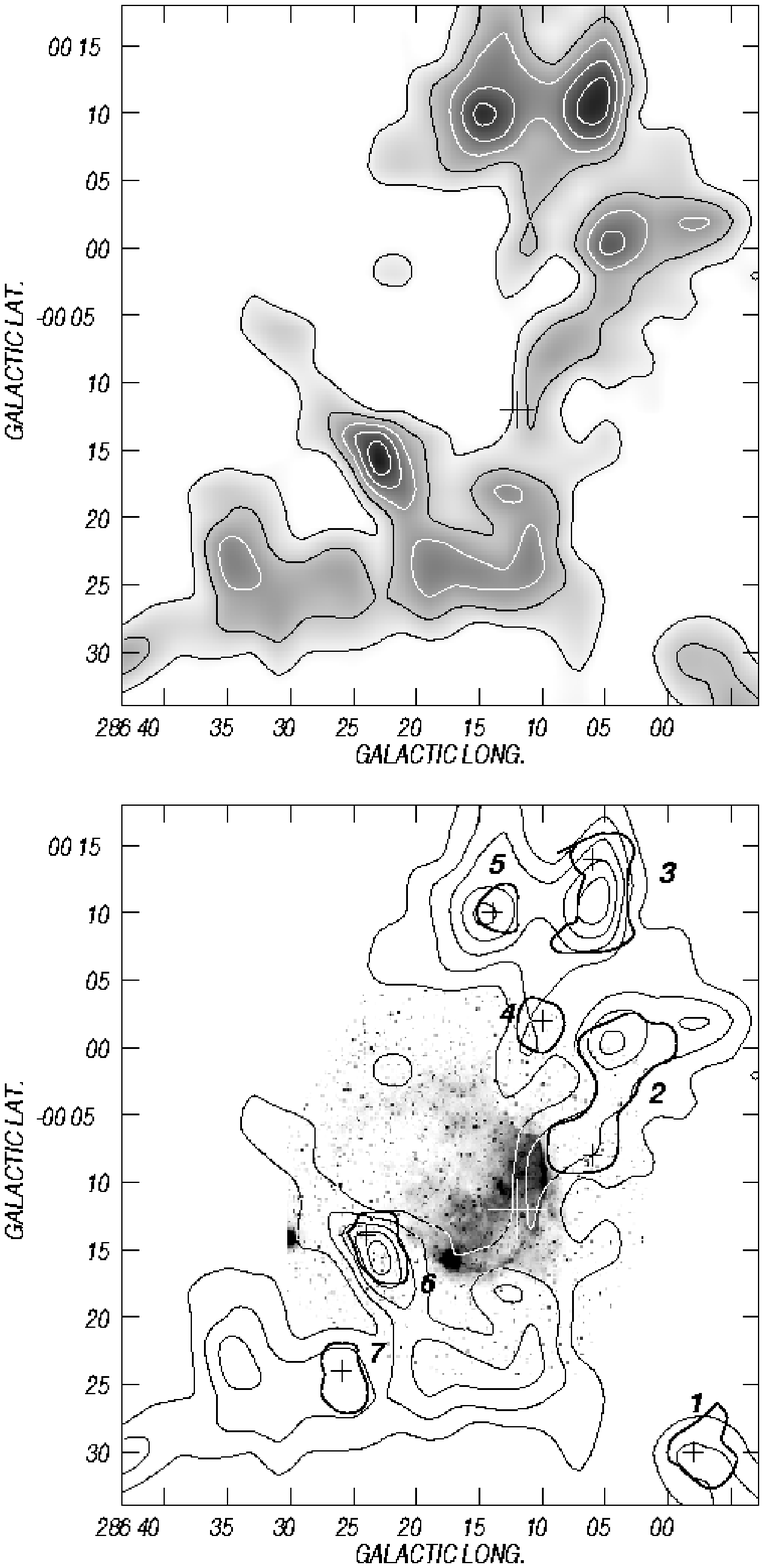}}
\caption{{\it Top:} $^{12}$CO(1-0) emission distribution corresponding 
to the velocity interval $-$27.2 to $-$14.0 \kms. The grayscale is from 
15 to 150 K \kms. 
The contour lines are 20.2, 40.5, 60.7, 81.0 and 101.2 K \kms. 
{\it Bottom:} Overlay of the SuperCOSMOS  ({\it grayscale)} and the CO 
({\it contours}) images. The thick lines delineate the C$^{18}$O cores
described by Yonekura et al. (2005). The numbers correspond to the 
C$^{18}$O cores as identified in Table 2 by Yonekura et al., and the crosses
mark the core positions indicated in the same table. 
}
\label{ir}
\end{figure}

\subsection{The emission in the infrared}

The distribution of the IR emission at 60 and 100 $\mu$m, due to thermal
dust emission, is displayed in Fig. 6. The upper panels show the emission 
at both  wavelengths while the bottom panels display overlays of the IR 
emission and the optical  image of Gum\,31. The images reveal an 
IR structure which is brighter near  {\it(l,b)} = 
(286\deg23\arcmin,--0\deg15\arcmin) and where the optical nebula has its 
sharpest border, and is weaker on the opposite side. The brightest IR 
emission regions at 60 $\mu$m\ and 100 $\mu$m\ are projected onto the 
neutral envelope, delineating  the ionized nebula.  The IR emission at
(286\deg23\arcmin,--0\deg15\arcmin) detected at both wavelengths coincides
with strong $^{12}$CO(1-0) emission. $^{12}$CO emission also appears 
bordering the two IR clumps located near the bright rim at 
{\it (l,b)} = (286\deg 10\arcmin, --0\deg 10\arcmin) detected at 60 $\mu$m,  
and in between. The distribution of the IR emission at both 
wavelengths shows the presence of dust most probably related 
to the surrounding \hi\  and molecular shells.

Following the procedure described by Cichowolski et al. (2001), we derived
the color temperature of the dust  associated with the \hii\ region and 
the neutral envelope based on the IR fluxes at 60 $\mu$m and 100 $\mu$m.
Taking into account different values for the background emission, we 
found $T_d$ =  34$\pm$7 K. The range of temperatures corresponds to  
$n$ = 1-2 and to different IR background emissions. The parameter $n$ is 
related to the dust absorption efficiency
($\kappa_\nu\ \propto\ \nu^n$). We adopted $\kappa_\nu$ = 40 
cm$^2$ g$^{-1}$. This value was derived from the expressions by Hildebrand
(1983) for $n$ = 1 and  $\lambda$ = 60 $\mu$m.
The dust temperature obtained is typical for \hii\ regions.

\begin{figure*}
\resizebox{12cm}{!}{\includegraphics{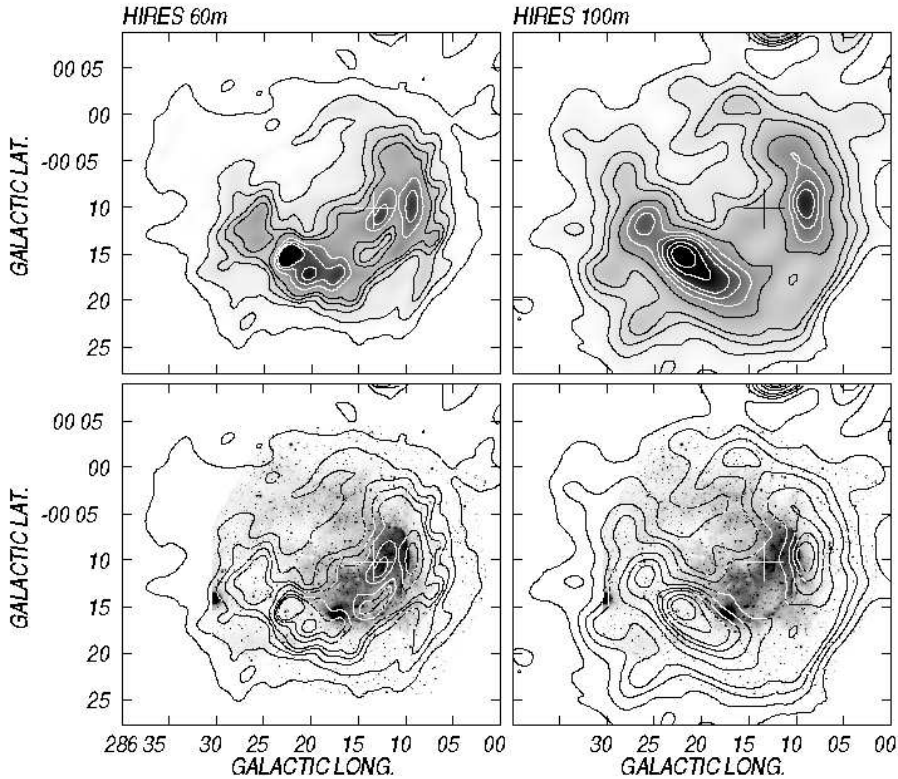}}
\caption{{\it Top:} Far infrared IRAS images at 60 and 100 $\mu$m towards
Gum\,31. The grayscale is from 300 to 3000 MJy ster$^{-1}$ for the image
at 60 $\mu$m, and from 600 to 4500 MJy ster$^{-1}$ for the image at 100 $\mu$m.
The contour lines are from 200 to 1000 MJy ster$^{-1}$ in steps of 200
MJy ster$^{-1}$ and from 1000 to 3000 MJy ster$^{-1}$ in steps of 500
MJy ster$^{-1}$ for the image at 60 $\mu$m, and from 400 to 1000 MJy 
ster$^{-1}$ in steps of 200 MJy ster$^{-1}$ and from 1000 to 3000 MJy 
ster$^{-1}$ in steps of 500 MJy ster$^{-1}$ for the image at 100 $\mu$m. 
{\it Bottom:} Overlay of the optical (SuperCOSMOS) ({\it grayscale)} and the 
IR images. 
}
\label{iras-ir}
\end{figure*}

Figure 7 shows an overlay of the distribution of the emission in the MSX 
bands A and E, and the optical image. The emission in band A closely follows 
the brightest sections of the nebula, and correlates with the neutral gas. 
Particularly, the brightest regions emitting at 8.3 $\mu$m coincide with 
the dense cores 2 and 6 (see Fig. 5) found by Yonekura et al. (2005). On 
the contrary, the emission in band E appears clearly associated with the 
ionized gas. 
Note that the strongest emission region in band A at {\it(l,b)} = 
(286\deg23\arcmin,--0\deg15\arcmin) coincides with bright molecular and far 
infrared clumps.

The emission distribution in band A is most probably related to 
emission from polycyclic aromatic hydrocarbons (PAHs). According to
Cesarsky et al. (1996), these dust grains cannot survive inside the \hii\ 
region, but can on the neutral PDR, where they radiate in the PAH bands
at 7.7 and 8.6 $\mu$m, included in the MSX band A. MSX band E, on the
contrary, includes continuum emission from very small grains, which 
can survive  inside ionized regions (cf. Deharveng et al. 2005), and 
a contribution from nebular emission lines.

In summary, the distribution of the  neutral atomic and molecular gas, 
and that of the interstellar dust reveals the presence of a neutral 
shell surrounding the \hii\ region. The distribution of the molecular gas 
and that of the emission in the MSX band A related to PAHs strongly suggests 
the presence of a PDR at the interface between the ionized and the molecular 
gas.  

Both the infrared emission at 60 and 100 $\mu$m and the optical image 
suggest that the faint optical emission region near  {\it (l,b)} = 
(286\deg 20\arcmin, --0\deg 4\arcmin) (indicated by an arrow in the 
bottom right image of Fig. 6) are also linked to Gum\,31.
As pointed out in Sect. 3.1, this region is also faint in the radio 
continuum (see Fig. 2), and corresponds to weak regions in the \hi\ and CO 
envelopes suggesting that the ionizing flux of the massive stars in the 
open cluster may drain through these regions to the general interstellar 
medium. The situation resembles the case of the stellar wind bubble around 
WR\,23 (Cappa et al. 2005).

\begin{figure*}
\resizebox{12cm}{!}{\includegraphics{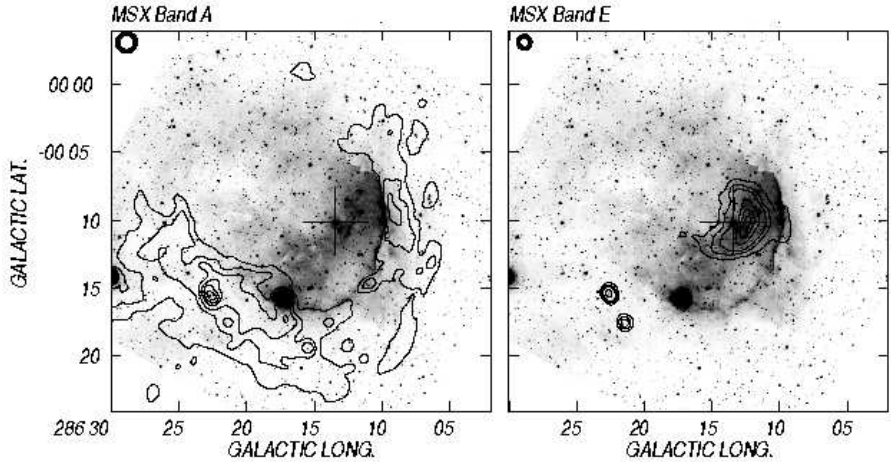}}
\caption{Overlay of the MSX infrared images (contours) corresponding to 
bands {\rm A} (8.28 $\mu$m) and {\rm E} (21.34 $\mu$m), 
and the SuperCOSMOS  image of the nebula (grayscale). The contour 
lines are 25, 39, 57, 85, 114, and 140 MJy ster$^{-1}$ for band A; 
and 36, 46, 57, 85, 114, and 140 MJy ster$^{-1}$ for band E.
 }
\label{msx-ir}
\end{figure*}

\section{Discussion}

\begin{table*}
\centering
\caption{Main parameters of the ionized and neutral gas in Gum\,31}
\begin{tabular}{lc}
\hline
Adopted distance (kpc)              &  3.0$\pm$0.5  \\
\hline
{\it \hii\ region} \\
Flux density at 4.85 GHz (Jy)       &  37.7$\pm$2.5  \\
Angular radius (arcmin)             &  7.5$\pm$0.2      \\
Linear radius (pc)                  &  6.5$\pm$1.0  \\
Emission measure (pc cm$^{-6}$)     &  (1.5$\pm$0.2)\por 10$^4$  \\
rms electron density [$n_e$](\cmtres)      &  33$\pm$3  \\
Used Lyman UV photons [$ log N_{Ly-c}$] (s$^{-1}$)  
                &  49.0$\pm$0.1  \\
Ionized mass (\msun)                & 3300$\pm$1100  \\
\hline
{\it Neutral atomic shell} \\
{\it (l,b)} centroid                & 286\deg 15\arcmin,--0\deg 10\arcmin  \\
Velocity interval v$_1$,v$_2$ (\kms)  & $-$13,$-$32  \\
\hi\ systemic velocity (\kms)       & $-$23        \\
Expansion velocity $v_{exp}$ (\kms) & 11           \\
Radius of the \hi\ structure (arcmin)  & 11.5       \\
Radius of the \hi\ structure $R$ (pc)  & 10.0$\pm$1.7 \\
Atomic mass in the shell (\msun)       &  1500$\pm$500         \\
\hline
{\it Molecular shell}\\
Velocity interval v$_1$,v$_2$ (\kms)  & --27.2,--14.0 \\
Mean H$_{2}$ column density (\cmdos) &  1.2\por 10$^{22}$ \\
Molecular mass (\msun)              &   (1.1$\pm$0.5)\por 10$^5$ \\
\hline
{\it Dust related to the \hii\ region and the neutral shell} \\
Total dust mass (\msun)                   & 60$\pm$20        \\
Dust color temperature (K)       & 34$\pm$7       \\
\hline
\end{tabular}
\end{table*}

\subsection{Physical parameters}

The main parameters of the dust and the ionized and neutral gas related 
to Gum\,31 are summarized in Table 2. The parameters of the ionized gas 
were derived from the image at 4.85 GHz. The uncertainty in the flux 
density corresponds  to an error of 0.1 \jyb\ in the estimate of the 
radio continuum background. The electron density and the \hii\ mass were 
obtained from the expressions by Mezger \& Henderson (1967) for a spherical 
\hii\ region of constant electron density (rms electron density $n_e$). 
The presence of He\,{\sc ii} was considered by multiplying the \hi\ mass by 
1.27. The number of UV photons necessary to ionize the gas 
$log \ N_{Ly-c}$ was derived from the radio continuum results. Errors in  
the linear radius, in the rms electron density, and in the excitation 
parameter come from the distance uncertainty. The high electron 
density of the \hii\ region is compatible with the relatively short 
lifetime for the cluster.
 
The parameters of the neutral atomic gas includes: the {\it (l,b)} position
of the centroid of the \hi\ shell, the velocity interval spanned by
the structure, the systemic and expansion velocities, the radius of the 
neutral gas structure, and the associated atomic mass.

The expansion velocity was estimated as in previous papers (see Cappa et al. 
2005 and references therein) as $v_{exp} = (v_2 - v_1$)/2 + 1.6 \kms. The 
extra 1.6 \kms\ allows for the presence of \hi\ in the caps of the expanding 
shell, which should be present in the images at v$_1$+1.6 km/s and 
v$_2$--1.6 km/s. These caps are not detected in the present case, as in most 
of the cases, probably because of their weak brightness temperature.  

The radius of the \hi\ structure was estimated from Fig. 4 and corresponds 
to the position of the maxima in the shell. The neutral atomic mass 
corresponds to the mass excess in the shell, assuming that the gas is 
optically thin and including a He abundance of 10\%.

The $H_2$ column density ($N_{H2}$) and the molecular mass  were estimated 
from the $^{12}$CO data, making use of the empirical relation between 
the integrated emission $W_{CO}$ (= $\int T dv)$ and $N_{H2}$. We 
adopted $N_{H2}$ = (1.06$\pm$0.14) \por W$_{CO}$ $\times$ 10$^{20}$ cm$^{-2}$ 
(K \kms)$^{-1}$, obtained from $\gamma$-ray studies of molecular clouds in 
the Orion region (Digel \etal\ 1995). To derive the molecular mass, 
we integrated the CO emission within a circle of about 16\arcmin\ in radius 
(= 15 pc at 3.0 kpc), centered at {\it(l,b)} = 
(286\deg16\arcmin,--0\deg8\arcmin). 

The ambient density derived by distributing the molecular mass over a 
sphere of 15.0 pc in radius is  $\simeq$350 \cmtres, reinforcing the idea 
that the molecular gas represents the remains of the original material
where the open cluster NGC\,3324 was born. The difference between 
the electron density and the ambient density indicates that 
the \hii\ region is expanding.

It is known that massive stars have stellar winds with high mass
loss rates and terminal velocities (Chlebovski \& Garmany 1991; Prinja
et al. 1990), which generally affect the surrounding interstellar medium 
creating {\it interstellar bubbles} (Weaver et al. 1977). In the optical 
domain, they appear as ring-shaped structures (Lozinskaya 1982). 
The optical appearance of the \hii\ region, almost spherical without  
clear evidence for a central cavity, suggests that the massive stars in 
the cluster have weak stellar winds (which is supported by the short 
age of the cluster) and/or have existed during a very short 
period of time to create an interstellar bubble in an interstellar medium
as dense as observed here.

\begin{figure}
\resizebox{8.0cm}{!}{\includegraphics{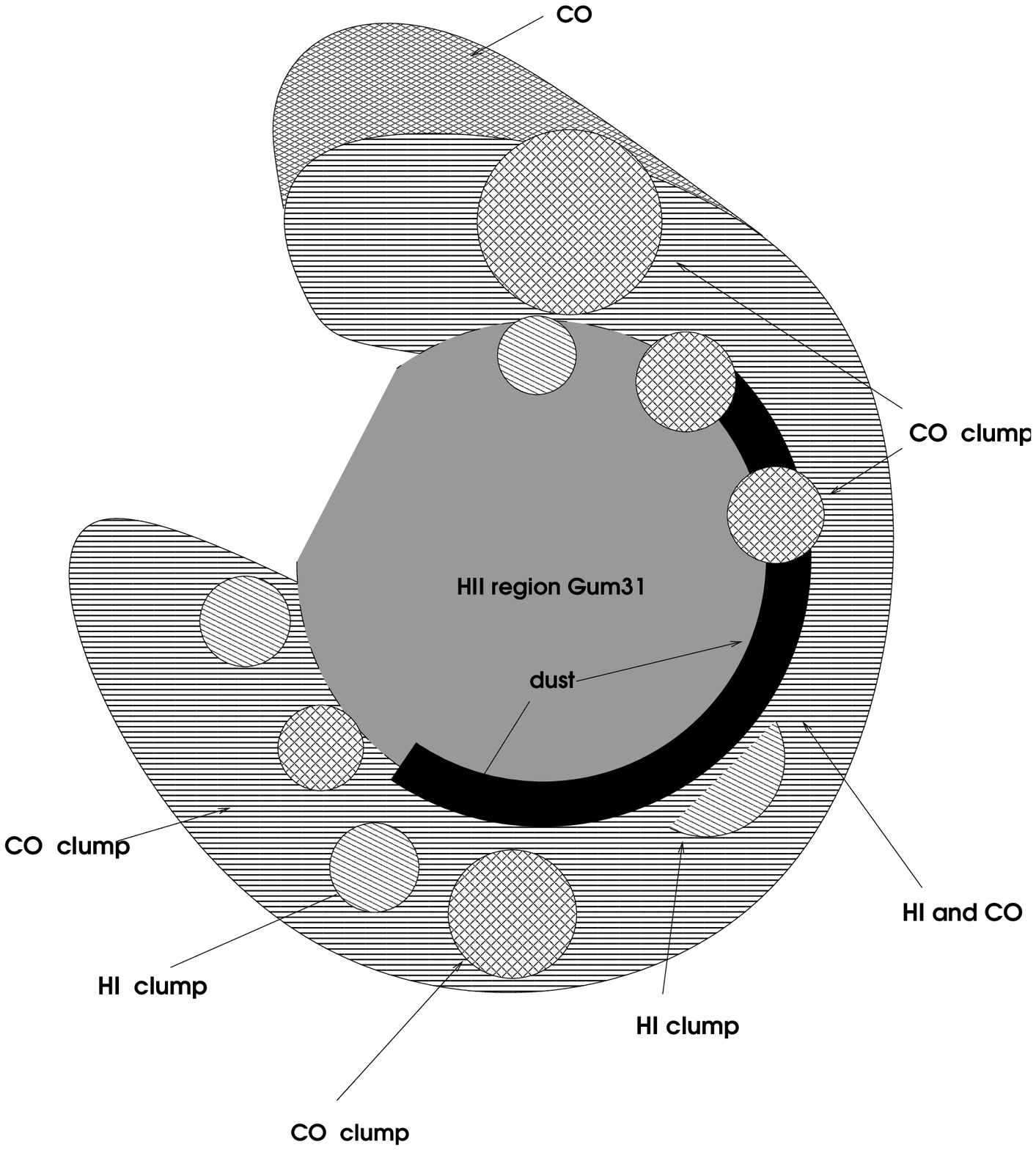}}
\caption{Schematic view of Gum\,31 and the respective locations of the
different gas components and of interstellar dust.}
\end{figure}

Figure 8 shows schematically the distribution of the different gas 
components. Although \hi\ and molecular clumps anticorrelate in position, 
the large scale \hi\ and molecular gas coincide in position around the 
nebula but in the region at higher galactic latitudes, where the CO 
emission is observed outside the \hi\ region.
The large amount of molecular gas compared to the neutral atomic gas
supports the idea that the CO emission which encompasses the ionized 
nebula represents the remains of the molecular material where the open 
cluster was born. Part of this \hi\ gas may have  originated in the
photodissociation of the molecular gas.

\begin{table*}
\centering
\caption[]{IR Point sources and YSO candidates from the IRAS, MSX and 2MASS
catalogs}
\begin{tabular}{rllcrrrrc}
\hline
\hline
 $\#$ &    $l$[$\circ\ \arcmin$]    &  $b$[$\circ\ \arcmin$]    & {\it IRAS} source  
&     \multicolumn{4}{c}{Fluxes[Jy]}  & $L_{IRAS}[10^3 L_{\sun}]$ \\
 & & &  &  12$\mu$m  &    25$\mu$m &   60$\mu$m &  100$\mu$m &  \\
\hline
1 &  285\deg 59\farcm 82 & +0\deg 05\farcm 82 & 10349-5801 &   1.6 &   2.5 &   27.9 &  104  &  1.5\\
2 &  286\deg 04\farcm 98 & --0\deg 25\farcm 08 & 10335-5830 &   0.8 &   3.5 &   22.8 & 111  &  1.5 \\
3 &  286\deg 08\farcm 58 & --0\deg 06\farcm 96 & 10351-5816 &   6.3 &   6.8 &  315   & 1480  & 18\\ 
4  &  286\deg 08\farcm 64 & --0\deg 18\farcm 78 & 10343-5826 &   5.5 &   4.3 &  103   & 340  &  5 \\ 
5  &  286\deg 10\farcm 92  & --0\deg 14\farcm 7 &  10349-5824 & 5.9 &  9.0 & 163 & 1660  &  18\\
6 &  286\deg 12\farcm 18 & +0\deg 10\farcm 20 & 10365-5803 &   7.2 &  86   & 1170   & 2780  & 43  \\
7 &  286\deg 15\farcm 12 & --0\deg 25\farcm 44 & 10346-5835 &   1.1 &   3.3 &   11.7 & 1430  &  14\\
8 &  286\deg 17\farcm 34 & +0\deg 00\farcm 42 & 10365-5814 &   2.3 &   1.7 &   85   & 246  &  4\\
9 &  286\deg 22\farcm 5 & --0\deg 15\farcm 3 & 10361-5830 &  12.4 &  38.4 &  626   & 2160  &  30\\
10 &  286\deg 26\farcm 94 & --0\deg 22\farcm 92 & 10361-5839 &   2.8 &   5.4 &   51.6 & 307   & 4 \\
11 &  286\deg 29\farcm 34 & --0\deg 16\farcm 68 & 10368-5835 &   2.1 &   2.3 &   84   & 271   & 4 \\
12 &  286\deg 33\farcm 72 & --0\deg 07\farcm 08 & 10379-5828 &   1.0 &   2.0 &   13.8 &  540   & 5 \\
\hline
\hline
$\#$ &   $l$[$\circ \arcmin$] &  $b$[$\circ\ \arcmin$] &   {\it MSX} source 
&     \multicolumn{4}{c}{Fluxes[Jy]} \\     
 &  &  &  &  8$\mu$m &  12$\mu$m & 14$\mu$m & 21$\mu$m & Class. \\
\hline
13 & 286\deg 09\farcm 78 & --0\deg 11\farcm 28 & G286.1626-00.1877 & 0.7311 & 1.298 & 1.256 & 2.806 &  C\hii\ \\
14 & 286\deg 12\farcm 36 & --0\deg 09\farcm 66 & G286.2056-00.1611 & 0.1585 & 0.9313 & 2.151 & 6.904  & MYSO\\
15 & 286\deg 12\farcm 48 & --0\deg 10\farcm 32 & G286.2077-00.1720 & 0.0870 & 0.9481 & 1.764 & 2.9 &  MYSO \\
16 & 286\deg 12\farcm 54 & +0\deg 10\farcm 14  & G286.2086+00.1694 & 1.353  & 2.882  & 7.182 & 40.57  & MYSO\\
17 & 286\deg 12\farcm 6  & --0\deg 10\farcm 68 & G286.2096-00.1775 & 0.2202 & 0.9538 & 1.368 & 7.206  & MYSO \\
18 & 286\deg 15\farcm 42 & --0\deg 19\farcm 44 & G286.2566-00.3236 & 2.04   & 2.151  & 1.34 & 4.328  & C\hii \\
19 & 286\deg 21\farcm 48 & --0\deg 17\farcm 58 & G286.3579-00.2933 & 0.7126 & 1.815 & 2.677 & 6.065  & MYSO\\
20 & 286\deg 22\farcm 5 & --0\deg 15\farcm 78 & G286.3747-00.2630 & 3.591  & 4.756 & 2.409 & 7.577  & C\hii\ \\
21 & 286\deg 22\farcm 62 & --0\deg 15\farcm 36 & G286.3773-00.2563 & 1.628 & 2.918 & 3.855 & 12.1  & MYSO \\
\hline
\hline
$\#$ &  $l$[$\circ \arcmin$]   &  $b$[$\circ \arcmin$]  &   {\it 2MASS} source  
& $J$[mag] & $H$[mag] & $K_s$[mag] & $(J-H)$ & $(H-K)$  \\
\hline
22 & 285\deg 54\farcm 42 & --0\deg 13\farcm 74 &  10350210-5831039 & 10.584 & 10.631 & 10.564 & -0.047 & 0.067\\
23 & 286\deg 04\farcm 44 & --0\deg 00\farcm 24 & 10365972-5824186 & 12.042 & 10.51  & 9.284  & 1.532  & 1.226\\
24 & 286\deg 06\farcm 36 & --0\deg 08\farcm 64 & 10364112-5832326 & 11.367 & 11.385 & 11.28  & -0.018 & 0.105\\
25 & 286\deg 06\farcm 54 & --0\deg 08\farcm 40 & 10364296-5832267 & 10.465 & 10.513 & 10.439 & -0.048 & 0.074\\
26 & 286\deg 09\farcm 60 & --0\deg 00\farcm 30 & 10373406-5826540 & 12.487 & 11.454 & 10.513 & 1.033  & 0.941\\
27 & 286\deg 09\farcm 78 & --0\deg 11\farcm 22 & 10365396-5836293 & 12.242 & 10.99  & 10.103 & 1.252  & 0.887\\
28 & 286\deg 10\farcm 08 & --0\deg 09\farcm 12 & 10370395-5834489 & 10.772 & 9.712  & 8.685  & 1.06   & 1.027\\
29 & 286\deg 10\farcm 20 & --0\deg 11\farcm 10 & 10365749-5836366 & 13.99  & 12.815 & 11.984 & 1.175  & 0.831\\
30 & 286\deg 11\farcm 16 & --0\deg 02\farcm 64 & 10373574-5829405 & 15.318 & 13.343 & 11.614 & 1.975  & 1.729\\
31 & 286\deg 12\farcm 96 & --0\deg 04\farcm 86 & 10373956-5832311 & 10.926 & 10.752 & 10.554 & 0.174  & 0.198\\
32 & 286\deg 13\farcm 02 & --0\deg 10\farcm 86 & 10371717-5837460 & 11.793 & 11.773 & 11.665 & 0.02   & 0.108\\
33 & 286\deg 13\farcm 32 & --0\deg 08\farcm 46 & 10372824-5835492 & 12.191 & 11.301 & 10.448 & 0.89   & 0.853\\
34 & 286\deg 13\farcm 38 & --0\deg 10\farcm 20 & 10372226-5837229 & 7.563  & 7.588  & 7.479  & -0.025 & 0.109\\
35 & 286\deg 13\farcm 92 & --0\deg 17\farcm 87 & 10365763-5844052 & 12.22  & 11.882 & 11.51  & 0.338  & 0.372\\
36 & 286\deg 19\farcm 26 & --0\deg 06\farcm 84 & 10381421-5837192 & 12.635 & 11.886 & 11.349 & 0.749  & 0.537\\
37 & 286\deg 20\farcm 10 & --0\deg 19\farcm 80 & 10373105-5849026 & 15.155 & 12.8   & 11.127 & 2.355  & 1.673\\
38 & 286\deg 21\farcm 06 & --0\deg 16\farcm 86 & 10375219-5847133 & 12.271 & 11.363 & 10.675 & 0.908  & 0.688\\
39 & 286\deg 21\farcm 90 & --0\deg 04\farcm 92 & 10383875-5836566 & 12.116 & 11.636 & 11.187 & 0.48   & 0.449\\
40 & 286\deg 22\farcm 92 & --0\deg 13\farcm 20 & 10381461-5844416 & 12.367 & 11.815 & 11.398 & 0.552  & 0.417\\
41 & 286\deg 24\farcm 92 & --0\deg 18\farcm 66 & 10380736-5850240 & 12.477 & 11.538 & 10.78  & 0.939  & 0.758\\
42 & 286\deg 28\farcm 11 & --0\deg 24\farcm 50 & 10380702-5857039 & 12.724 & 11.599 & 10.726 & 1.125  & 0.873 \\
43 & 286\deg 28\farcm 44 & --0\deg 23\farcm 71 & 10381226-5856318 & 13.348 & 12.274 & 11.355 & 1.074  & 0.919\\
44 & 286\deg 29\farcm 46 & --0\deg 06\farcm 30 & 10392451-5841486 & 13.836 & 12.561 & 11.619 & 1.275  & 0.942\\
45 & 286\deg 29\farcm 46 & --0\deg 08\farcm 10 & 10391799-5843257 & 13.195 & 12.322 & 11.624 & 0.873  & 0.698\\
46 & 286\deg 30\farcm 90 & --0\deg 06\farcm 30 & 10393410-5842321 & 11.701 & 10.423 & 9.334  & 1.278  & 1.089\\
47 & 286\deg 31\farcm 27 & --0\deg 28\farcm 40 & 10381363-5902003 & 13.368 & 12.557 & 11.982 & 0.811  & 0.575\\
\hline
\hline
\end{tabular}
\end{table*}

\subsection{Energy budget}

Bearing in mind that massive stars have a copious UV flux capable of
ionizing the surrounding \hi\ gas, we investigate in this section whether 
the massive stars in NGC\,3324 can provide the energy to ionize the gas.

As described in Sect. 1, the only O-type stars in the open cluster are
the multiple system HD\,92206, which contains three O-type stars classified 
as O6.5V, O6.5V and O8.5 (Walborn 1982, Mathys 1988). Considering that
HD\,92206A is almost 1 mag brighter than component B (Clari\'{a} 1977), the 
similar spectral types found by Mathys (1988) suggest that HD\,92206A
is probably a spectroscopic binary O6.5V + O6.5V. Taking into 
account the UV photon fluxes emitted by the stars, $N_{*}$ (s$^{-1}$), 
derived by Martins et al. (2005) from stellar atmosphere models, a group of 
four massive stars having the spectral types indicated above have a  total 
UV photon flux corresponding to $log \ N_*$ =  49.4. By comparing $N_{*}$ 
with the UV photons used to ionize the gas ( $N_{Ly-c}$)  listed in Table 2, 
we conclude that the massive O-type stars in NGC\,3324 can maintain the 
\hii\ region ionized. 
 
The radius of the Str\"omgren sphere formed in a region with a volumetric
ambient density of 350  cm$^{-3}$ is $\simeq$ 1.7 pc, which is lower than the 
radius of the ionized region. Following Spitzer (1978), we estimated that
the \hii\ region has been expanding for $\approx$ 1\por 10$^6$ yr.

The ambient density we  derived from the  $^{12}$CO(1-0) data is larger than 
the rms electron density, 33 cm$^{-3}$ (see Table 2), giving additional 
support for the interpretation that the \hii\ region is expanding. 

\begin{figure*}
\resizebox{13cm}{!}{\includegraphics{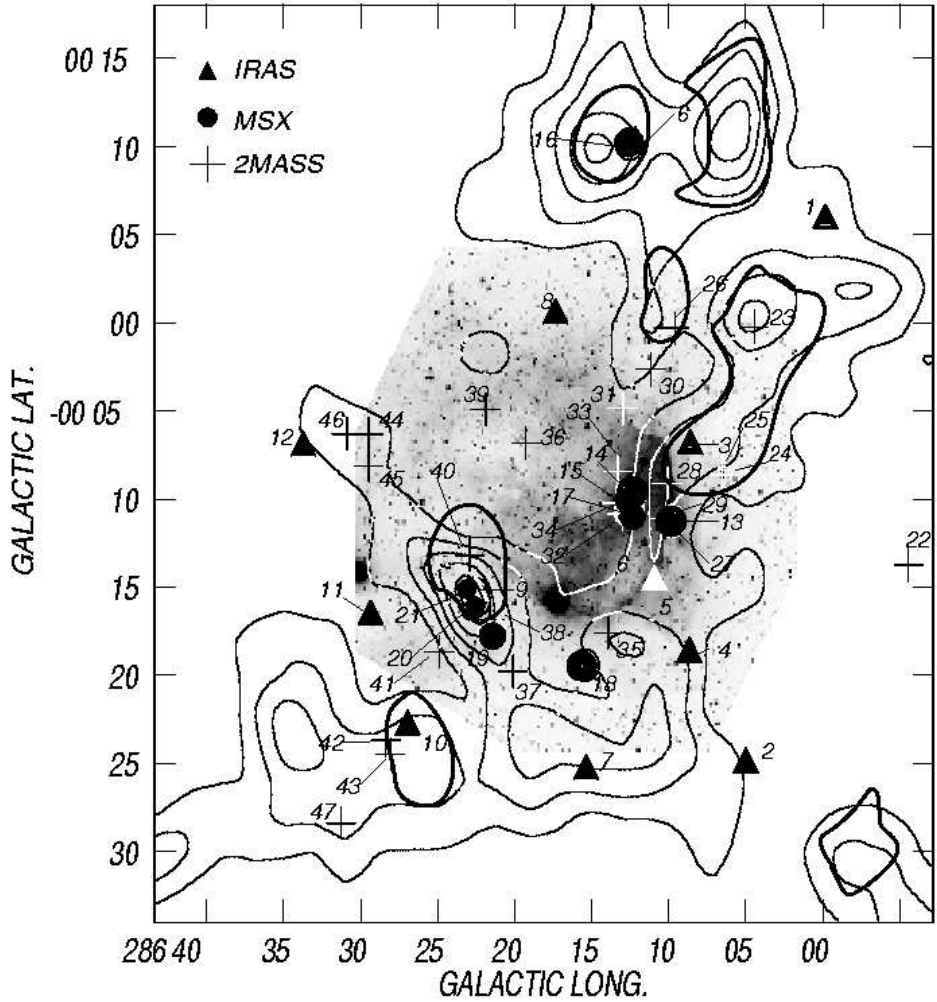}}
\caption{{\it Top:} Point sources from the IRAS (triangles), MSX (circles),
and 2MASS catalogs overlaid onto the SuperCOSMOS image and the
$^{12}$CO and C$^{18}$O contours (see Fig. 5 for details).
 }
\label{ir}
\end{figure*}

\section{Stellar formation}

Stellar formation may be induced by the expansion of the \hii\ region in 
the surrounding molecular envelope, where the presence of high density 
regions favors the conditions which lead to the formation of new stars. 
Among the different processes that induce star formation, the {\it collect 
and collapse} process proposed by Elmegreen \& Lada (1977)  may work 
efficiently in the dense molecular shells around \hii\ regions. 
The physical conditions in the molecular shells were discussed 
by different authors (Whitworth et al. 1994; Hosokawa \& Inutsuka 2005; 
2006, Dale et al. 2007).

To investigate the presence of protostellar candidates in this region 
we used data from the IRAS, MSX and 2MASS point source catalogs.
We searched for point sources in a region of  about 20\arcmin\ in radius 
centered at the position of the  NGC\,3324.

\subsection{IRAS sources}

The IRAS point source catalog allows identification of protostellar
candidates following the criteria by Junkes et al. (1992). Sources 
with quality factors $Q_{60} + Q_{100} \geq$ 4 were considered.  Twelve out 
of the thirteen point sources have IR spectra compatible with 
protostellar objects. 
The names of the  IRAS  protostellar candidates, their {\it (l,b)} 
position, their fluxes at different IR wavelengths, and their luminosity
derived following Yamaguchi et al. (2001), along with a reference number 
are listed in Table 3. 
The IRAS protostellar candidates are indicated as filled triangles in Fig. 
9, superimposed onto the molecular gas and ionized gas distributions. 
Each source can be identified in the figure by its reference number.
They are probably young stellar objects (YSOs).

\subsection{MSX sources}

Massive young stellar objects (MYSOs) can be identified from the MSX point
catalogue following the criteria by Lumsden et al. (2002). A total of  
310 MSX point sources were found to be projected onto the 
area. 

Lumsden et al. (2002) found that MYSOs have infrared fluxes with ratios
$F_{21}/F_8 >$ 2 and $F_{14}/F_{12} >$ 1, where $F_{8}$, $F_{12}$, 
$F_{14}$, and $F_{21}$ are the fluxes at 8.3, 12, 14, and 21 $\mu$m.
For compact \hii\ regions, ratios are $F_{21}/F_8 >$ 2 and 
$F_{14}/F_{12} <$ 1.
Taking into account sources with flux quality q $\geq$ 2, we were 
left with 14 sources after applying Lumsden et al.'s criteria
indicated above.
 
Six out of the 14 sources can be classified as MYSOs, while three sources 
are compact \hii\ regions, also indicative of active stellar formation.
The nine sources are listed in Table 3 and indicated in Fig. 9 as circles.

\subsection{2MASS sources}

Point sources with infrared excess, which are YSO candidates, were searched 
for in the 2MASS catalog (Cutri et al. 2003), which provides detections 
in three near IR bands: $J$, $H$, and $K_S$, at 1.25, 1.65, and 2.17 $\mu$m, 
respectively. A total of 2$\times$10$^4$  sources are projected onto 
a circular region of 20\arcmin\ in radius. We took into account sources 
with $S/N >$ 10 (corresponding to quality ``AAA''). 
Only sources with  $K_S <$ 12 were included. The last criterium corresponds 
to stars with spectral types earlier than B3 at a distance of 3.0 kpc. 
Following Comer\'{o}n et al. (2005) and Romero (2006), we determined the  
parameter $q$ = $(J-H) - 1.83 \times\ (H-K_s)$. Sources with $q \leq$ --0.15 
are classified as objects with infrared excess which may be YSOs. After 
applying this criterium we were left with 26 sources, which are shown in 
the color-color and magnitude-color diagrams of Fig. 10. Absolute magnitudes 
for the ZAMS in the K$_S$ band are from Hanson et al. (1997), while values 
corresponding to {\it (J-H)} and {\it (H-K$_S$} colors were obtained from 
Koornneef (1983). The location of these sources is also marked in Fig. 9 
as crosses. The main data of these sources are shown in Table 3.

\begin{figure*}
\resizebox{17cm}{!}{\includegraphics{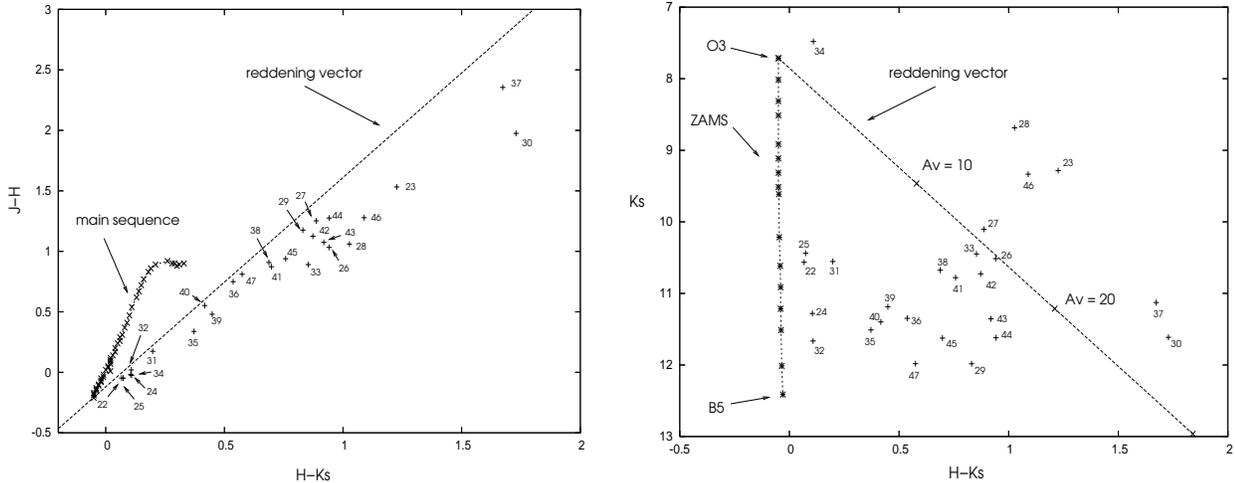}}
\caption{ Color-color and magnitude-color diagrams showing the 
2MASS sources with infrared excess.
 }
\label{ir}
\end{figure*}

The selected sources are located to the right of the reddening vector in 
the color-color diagram, where YSOs are expected to be placed. Most of the 
sources with the highest infrared excess appear projected onto the 
molecular clouds (see Fig. 9). The magnitude-color diagram shows that most 
of the sources are MYSOs. 
Although these  diagrams are not conclusive in identifying YSOs,
the strong infrared excess of the sources is compatible with
protostellar candidates.

\subsection{Distribution and characteristics of the YSOs}

Figure 9 shows that the IRAS and  MSX point sources, and most of the 2MASS 
point sources classified as YSOs appear bordering the ionized region, 
projected onto the molecular envelope detected in $^{12}$CO emission, close 
to the periphery of the \hii\ region,  or near the central cluster. 
Some are also coincident with the dense cores found by Yonekura
et al. (2005) in C$^{18}$O. We  analyze some particular regions. 

The IRAS source \#9, the MSX  sources \#19, \#20, and \#21, and the 2MASS 
sources \#38 and \#40, are projected onto a $^{12}$CO clump and onto 
the dense core 6 found in  C$^{18}$O. 
The presence of these sources indicates that stellar formation is ongoing  
in this particular molecular clump. As suggested  by Yonekura et al., a star 
cluster including massive stars is probably being formed in this region.
The MSX source \#19 is coincident with an extended radio source at 
843 MHz (see Fig. 2) and with a bright source in the MSX band E (see Fig. 7), 
showing that a compact \hii\ region may be present inside the molecular cloud. 
The MSX sources \#20 and \#21 coincide with a small source detected in 
both bands A and E (see Fig. 7) and with a point radio  source detected 
at 843 MHz (see Fig. 2). The detection of radio continuum emission is 
compatible with the classification of compact \hii\ region of source \#20. 
The fact that IR emission was detected in both bands A and E suggests 
contributions of warm interstellar dust and/or ionized gas. The radio
source coincident with source \#19 is about 48\arcsec \por 38\arcsec\
in size and its flux density is about 30 mJy. No electron density was
estimated since, very probably, the compact \hii\ region is optically thick 
at this low frequency. These results reinforce the idea that this
is a very active star forming region.

Particularly interesting is the bright rim region at {\it (l,b)} $\simeq$
(286\deg 10\arcmin,--0\deg 10\arcmin). The 2MASS sources \#27, \#28, and 
\#29, and the MSX source \#13 are located in this area, projected onto
the molecular envelope. IRAS sources \#3 and \#5 are also placed close
to the border of the \hii\ region, coincident with molecular material. 
Source \#3 coincides with a region emitting in the MSX band A and with
core 2. 
  
A bunch of protostellar objects is almost coincident in position with the open
cluster NGC\,3324: three MYSOs (MSX sources \#14, \#15, and \#17), and 
the 2MASS sources \#32, \#33, and \#34. The color-color diagram shows that 
two of these sources have relatively small IR excess. Some of these sources 
coincide with the loose IR cluster IC\,2599 listed by Dutra et al.(2003). 
These facts suggest that stellar formation is ongoing in the region of  
NGC\,3324,  as previously found by Carraro et al. (2001).
Also IRAS sources  \#6 and \#10 are projected onto the dense cores 5 and 7, 
respectively.
The large luminosities $L_{IRAS}$ estimated for some of these sources 
suggest that they are protostellar candidates for massive stars or star 
clusters.

The sources IRAS\,10355-5828 (286\deg 16\farcm 92,--0\deg 15\farcm 6) 
and MSX G286.2868-00.2604 (neither included  in Table 3 nor shown in Fig. 9) 
coincide with HD\,92207, a red supergiant whose membership to the open 
cluster is a matter of debate (Carraro et al. 2001). Its MSX fluxes 
correspond to an evolved object.

The detection of protostellar candidates in the IRAS, MSX, and 2MASS 
databases, strongly indicates that active stellar formation is currently 
ongoing in the  molecular shell around Gum\,31. 

To sum up, most of the  protostellar candidates detected towards
Gum\,31 appear projected onto the shell detected in the $^{12}$CO line, and
coincide with the dense cores detected in C$^{18}$O emission. Some
are located close to the bright sharp borders of the \hii\ region, and
near the open cluster NGC\,3324. 

The presence of protostellar objects on the molecular
envelope bordering the ionized region indicates that star formation
has been triggered by the expansion of the \hii\ region. The distribution
of the molecular and \hi\ gas around the ionized region suggests
that star formation could be due to the collect and collapse process.

\section{Summary}

We have analyzed the interstellar medium in the environs of the \hii\
region Gum\,31 to investigate the action of the massive stars in the exciting
open cluster NGC\,3324 on the surrounding neutral material. 

We based our study on \hi\ 21-cm line emission data belonging to the SGPS,
radio continuum data at  0.843, 2.4 and 4.85 GHz from the PMN Southern 
Radio Survey, $^{12}$CO data from Yonekura et al. (2005), and IRAS (HIRES) 
and MSX data.

Adopting a distance of 3.0$\pm$0.5 kpc, we have derived an ionized mass 
of 3300$\pm$1100 \msun\ and an electron density of 33$\pm$3 \cmtres. 
The four O-type stars in the HD\,92206 multiple system can provide the 
necessary UV photon flux to maintain the \hii\ region ionized. 

The \hi\ emission distribution in the environs of Gum\,31 shows the 
presence of an \hi\ shell approximately centered at the position of 
the multiple system HD\,92206. The \hi\ shell closely encircles the optical 
nebula. It is detected within the velocity range --32 to --13 \kms\ and
its systemic velocity of  --23 \kms\ is coincident, within errors, with the
velocity of the ionized gas in the nebula (--18 \kms). The \hi\ structure 
is 10.0$\pm$1.7 pc in radius and expands at about 11 \kms. The 
associated atomic mass is 1500$\pm$500 \msun.

Molecular gas with velocities in the range --27.2 and --14.0 \kms\  
surrounds the brightest parts of Gum\,31. The sharp interface
between the ionized and molecular material indicates that these gas
components are interacting. We have estimated a molecular
mass of  (1.1$\pm$0.5)\por 10$^5$ \msun. Two possible origins can
be suggested for the molecular shell: 1) it could have originated in 
the {\it collect and collapse} process driven by the expansion of the 
\hii\ region; or 2) it could represent the remains of the natal cloud 
where the open cluster NGC\,3324 formed. In both cases, a large amount of the 
neutral hydrogen in the \hi\ shell may have originated in the 
photodissociation of the molecular gas. 

The volume density of the molecular cloud, higher than the electron density 
(33 \cmtres), implies that the \hii\ region is expanding.

The emission in the far infrared correlates with  the \hii\ region and
the \hi\ envelope, indicating that the observed emission is probably
related  to the neutral and molecular envelopes.

The distribution of the emission in the MSX band A, which closely delineates 
the \hii\ region and correlates with the molecular emission, suggests the
presence of a PDR at the interface between the ionized and molecular gas.
 
A number of MSX, IRAS, and 2MASS point sources with IR spectra compatible
with protostellar objects appear projected onto the molecular envelope, 
implying that stellar formation is active in the higher density cores
of the molecular envelope around Gum\,31, where massive stars or star
clusters are probably being formed.

The optical image of the nebula does not show clear evidence of a central 
cavity, as expected in a stellar wind bubble, suggesting that the massive 
stars in the cluster have weak stellar winds or existed for a 
short period of time to develop an interstellar bubble in a high density
interstellar medium.

\begin{acknowledgements}
We dedicate this paper to our dear friend and collaborator Prof. Virpi 
Niemela, who passed away on December 18th, 2006.
C.E.C. is extremely grateful to her for her teaching and encouragement, and 
mainly for years of friendship. We thank the referee 
for many important comments and 
suggestions which largely improved this presentation. We also thank Dr. Y. 
Yonekura for making his CO data available to us.
This project was partially financed by the
Consejo Nacional de Investigaciones Cient\'{\i}ficas y T\'ecnicas (CONICET)
of Argentina under project PIP 5886/05 and PIP 5697/05, Agencia PICT 14018, 
and UNLP under projects 11/G072 and 11/G087. 
The Digitized Sky Survey (DSS) was produced at the Space Telescope Science
Institute under US Government grant NAGW-2166.

\end{acknowledgements}

\end{document}